\documentclass{article}
\usepackage{spconf,amsmath,epsfig}

\usepackage{amsmath,amsfonts}
\DeclareMathOperator{\sgn}{sgn}
\DeclareMathOperator*{\argmin}{arg\,min}
\usepackage{algorithm2e}

\newcommand{\comment}[1]{}

\title{
	Quantum-Assisted Greedy Algorithms
}
\name{
	Ramin Ayanzadeh$^{1,2*}$, John Dorband$^2$, Milton Halem$^2$ and Tim Finin$^2$
	\thanks{This research was supported by NASA grant (\#NNH16ZDA001N-AIST 16-0091), NIH-NIGMS Initiative for Maximizing Student Development Grant (2 R25-GM55036), and a Google Lime scholarship. We thank the D-Wave Systems management team, namely Rene Copeland, for granting us access to the D-Wave 2000Q quantum annealer.}
}

\address{
	$^1$Georgia Institute of Technology, Atlanta, GA 30332, US\\
	$^2$University of Maryland, Baltimore County, Baltimore, MD 21250, US
}
\begin{document}
%\ninept
%
\maketitle
\begin{abstract}
We show how to leverage quantum annealers (QAs) to better select candidates in greedy algorithms. 
Unlike conventional greedy algorithms that employ problem-specific heuristics for making locally optimal choices at each stage, we use QAs that sample from the ground state of a problem-dependent Hamiltonians at cryogenic temperatures and use retrieved samples to estimate the probability distribution of problem variables. 
More specifically, we look at each spin of the Ising model as a random variable and contract all problem variables whose corresponding uncertainties are negligible. 
Our empirical results on a D-Wave 2000Q quantum processor demonstrate that the proposed {\em quantum-assisted greedy algorithm ({\em QAGA})} scheme can find notably better solutions compared to the state-of-the-art techniques in the realm of quantum annealing.

\end{abstract}

\section{Introduction} 
Artificial intelligence (AI) is a disruptive technology that has transformed many industries. 
Quantum computing (QC) is a transformative computing phenomenon that promises to better address challenging problems that are intractable in the realm of classical computing. 
Quantum artificial intelligence (QAI) and quantum machine learning (QML) appear at the intersection of machine intelligence and QC, and aim to revolutionize application domains that are challenging for classical AI. 
Although promising, near-term quantum machines are susceptible to various sources of errors (such as decoherence and gate errors), and they do not include enough quantum bits (qubits) to accommodate quantum error correction policies. 
We are in the noisy intermediate-scale quantum (NISQ) era where we must perform computations in the presence of noise and infer the desirable output from erroneous outcomes~\cite{preskill2018quantum,huang2021information}. 

There are several models for the physical realization of quantum computers such as gate/circuit model, adiabatic, and measurement-based QCs. 
For the case of gate model QC, variational quantum approaches (a.k.a. classical-quantum hybrid schemes) are the leading candidate for demonstrating the supremacy of QAI/QML models. 
Conversely, the limited programmability of adiabatic quantum processors---i.e., adiabatic quantum computers and quantum annealers are single-instruction (quantum) computing machines that can only sample from the ground state of a given Hamiltonian---casts doubt on their ability to outperform classical AI/ML models~\cite{nielsen2010quantum}.  
In this paper, we introduce a novel variational scheme for quantum annealers (QAs) and demonstrate that it can outperform QAs in finding the ground state of Ising Hamiltonians.

Greedy algorithms are a problem-solving paradigm that makes locally optimal choices at each stage and expects them to yield a globally optimum solution. 
Although most greedy algorithms fail to achieve the global optimum,
they are often the best choice due to their efficiency~\cite{devore1996some}. 
As an example, greedy algorithms are widely used in sparse recovery applications at the cost of lower recovery accuracy (compared to convex optimization methods in compressive sensing~\cite{mousavi2019survey}. 
Note that greedy algorithms can find globally optimum solutions if the problem exhibits optimal substructure.

We introduce a novel hybrid approach, called {\em quantum-assisted greedy algorithms ({\em QAGA})}, that leverages QAs to better select candidates in each stage of a greedy algorithm. 
At each stage, QAGA employs a QA to provide samples from the ground state of the problem and use these retrieved samples to estimate the probability distribution of problem variables. 
After fixing variables with negligible uncertainties, QAGA proceeds to the next stage, where the QA will solve a smaller problem with sparser couplings. 
Our experimental results using a D-Wave 2000Q quantum processor show that QAGA can find samples with remarkably lower energy values compared to the best-known enhancements in the realm of quantum annealing.

\section{QAGA: A Hybrid Approach}
Quantum annealing is a meta-heuristic for addressing combinatorial (or discrete) optimization problems that are intractable in the realm of classical computing. 
\emph{Quantum annealers (QAs)} are a physical realization of the quantum annealing process that can draw samples from the ground state of the given Hamiltonians at cryogenic temperatures.  
The quantum processing unit (QPU) by D-Wave Systems is a single-instruction quantum computing machinery that can only sample from the ground state of the following Ising Hamiltonian: 
\begin{equation}	
	\label{eqn:ising_energy}
	\mathcal{H} := E_{\text{Ising}} \left( \mathbf{z} \right) = \sum_{i=1}^{N}{\mathbf{h}_i \mathbf{z}_i} + \sum_{i=1}^{N}{\sum_{j=i+1}^{N}{J_{ij} \mathbf{z}_i \mathbf{z}_j}},
\end{equation}
where $N$ denotes the number of quantum bits (qubits), spin variables $\mathbf{z} \in \{-1,+1\}^N$, and $\mathbf{h}$ and ${J}$ represent local fields  and couplers, respectively \cite{johnson2011quantum,mcgeoch2020theory,ayanzadeh2020leveraging}.

To solve a problem using QAs, we must define coefficients of the Ising Hamiltonian---i.e., $\mathbf{h}$ and ${J}$---such that the ground state of the corresponding $\mathcal{H}$ represents a solution of the problem of interest~\cite{ayanzadeh2020reinforcement}.
However, physical QAs are susceptible to various error sources, and therefore executing a quantum machine instruction (QMI) on a physical QA is not guaranteed to achieve the ground state of the corresponding $\mathcal{H}$~\cite{ayanzadeh2021multi}. 
Recent studies have shown that applying preprocessing and postprocessing policies can improve the fidelity of QAs. 
Nevertheless, such policies cannot bypass the technological barriers of physical QAs, and addressing hardware drawbacks can require device-level enhancements that may span generations of QAs~\cite{ayanzadeh2021multi}. 

We present the notion of {\em quantum-assisted greedy algorithms ({\em{QAGA})}} to improve the fidelity of the near-term QAs. 
Algorithm \ref{algo:qaga} illustrates the QAGA process. 
Let $\mathbf{z}^*$ be the ground state of $\mathcal{H}$---i.e., $\mathbf{z}^* = \argmin_{\mathbf{z}}{\mathcal{H}}.$
Our objective is to find a sample whose corresponding Ising energy value, shown in Eq.~\eqref{eqn:ising_energy}, approaches the energy value of the ground state of the given Ising Hamiltonian. 
More specifically, we aim to find $\tilde{\mathbf{z}}$ such that:
\begin{equation*}
	\label{eqn:ising_semi_optimality}
	\left| \mathcal{H}_{\mathbf{z}^*} - \mathcal{H}_{\tilde{\mathbf{z}}} \right| \to 0.
\end{equation*}

QAGA starts with $\tilde{\mathbf{z}}=\{\}$ and $\mathcal{H}^{t=0} = \mathcal{H}.$ 
At each iteration, QAGA uses a QA to draw ${n}$ samples from the ground state of $\mathcal{H}^t$. 
Let ${Z}$ denotes the set of all samples, drawn by a QA from the ground state of $\mathcal{H}^t$, as $Z=\{\mathbf{z}^1, \mathbf{z}^2, \cdots, \mathbf{z}^n \}$ where $\mathbf{z}^j \in \{-1,+1\}^N$. 
Every sample $\mathbf{z}^j$ contains a measurement for all qubits. 
Hence, we can look at each problem variable $\mathbf{z}_i$ as a random variable with Bernoulli distribution that takes its value from $\{-1, +1\}$. 
After retrieving the sample set ${Z}$, we estimate the uncertainty of every problem variable $\mathbf{z}_i$ as follows:
\begin{equation}
	\label{eqn:qaga_var_uncertainty}
	u(\mathbf{z}_i) = 1-\frac{|\sum_{j=1}^{n}{\mathbf{z}_i^j}|}{n}.
\end{equation}

For every variable $\mathbf{z}_i$ of $\mathcal{H}^t$ that $u(\mathbf{z}_i) \leq \theta,$ where $\theta \in [0,0.5)$ specifies the threshold parameter, we fix the value of optimum solution as: 
\begin{equation}
	\label{eqn:qaga_update_z*}
	\tilde{\mathbf{z}}_i = \sgn{ \left({ \sum_{j=1}^{n}{\mathbf{z}_i^j}} \right) }.
\end{equation}
since $\theta < 0.5$, $\tilde{\mathbf{z}_i}$ is guaranteed to take its value from $\{-1,+1\}$.

After fixing the value of $\mathbf{z}_i,$ we: (a) remove $\mathbf{h}_i$ from $\mathcal{H}^{t+1}$; (b) add the value of $\tilde{\mathbf{z}}_i J_{ij}$ (or $\tilde{\mathbf{z}}_i  J_{ji}$) to $\mathbf{h}_j$, for $j=1,2, \dots, N$ and $i \neq j$; and (c) remove the coupler $J_{ij}$ (or $J_{ji}$) from $\mathcal{H}^{t+1}$.
QAGA terminates when $\mathcal{H}^{t} \equiv \mathcal{H}^{t+1}$. 
If QAGA ends without fixing all problem variables, we apply the multi-qubit correction (MQC) method~\cite{ayanzadeh2021multi} on the samples from the last QAGA stage and assign values for the remaining problem variables. 

Contracting variables with negligible uncertainties results in a new Ising Hamiltonian ($\mathcal{H}^{t+1}$), which is smaller and sparser compared to $\mathcal{H}^{t}$. 
Hence, at each iteration of QAGA, the remaining Ising Hamiltonian becomes easier to solve with physical QAs.
Note that contracting variables reduces the number of qubits (${N}$) correspondingly. 
Finally, we can apply a classical local optimization heuristic, such as the single-qubit correction (SQC) method~\cite{ayanzadeh2021multi}, to increase the probability of finding the ground state.  

\begin{algorithm}[ht]
	\DontPrintSemicolon % Some LaTeX compilers require you to use \dontprintsemicolon instead
	\KwIn{$\mathcal{H}, \theta$}
	\KwOut{$\tilde{\mathbf{z}}$}
	$\tilde{\mathbf{z}} \gets \{\}$\;
	$\mathcal{H}^{t} \gets \mathcal{H}$\;
	$\mathcal{H}^{t+1} \gets \{\}$\;
	\While{$\mathcal{H}^{t+1} \ne \mathcal{H}^{t}$} {
		$\mathcal{H}^{t+1} \gets \mathcal{H}^{t}$\;
		$Z\gets\{\mathbf{z}^1, \mathbf{z}^2, \cdots, \mathbf{z}^n\} = \argmin_{\mathbf{z}} \mathcal{H}^{t}$\;
		\For{$i\gets0$ \KwTo ${N}$}{
			\If{$u(\mathbf{z}_i) \leq \theta$}{
				$\tilde{\mathbf{z}}_i \gets \sgn{ \left({ \sum_{j=1}^{n}{\mathbf{z}_i^j}} \right) }$\;
				$\mathcal{H}_{\mathbf{h}_i}^{t+1} \gets \{\}$\;
				\For{$j\gets0$ \KwTo ${N}$}{
					\If{$J_{ij} \in \mathcal{H}^{t+1}$}{
						$\mathcal{H}_{\mathbf{h}_j}^{t+1} \gets \mathcal{H}_{\mathbf{h}_j}^{t+1} + J_{ij}\tilde{\mathbf{z}}_i$\;
					$\mathcal{H}_{J_{ij}}^{t+1}  \gets \{\}$\;
					}
					\If{$J_{ji} \in \mathcal{H}^{t+1}$}{
						$\mathcal{H}_{\mathbf{h}_j}^{t+1} \gets \mathcal{H}_{\mathbf{h}_j}^{t+1} + J_{ji}\tilde{\mathbf{z}}_i$\;
					$\mathcal{H}_{J_{ji}}^{t+1}  \gets \{\}$\;
					}
				}

			}
	  }
	}
\If{$|\tilde{\mathbf{z}}| < N$}{
	$\hat{\mathbf{z}} \gets \mathrm{MQC}(Z)$\;
	$\tilde{\mathbf{z}} \gets \hat{\mathbf{z}} \cup \tilde{\mathbf{z}}$\;
}
	\Return{$\tilde{\mathbf{z}}$}\;
	\ \\
	\caption{Quantum-assisted greedy algorithm (QAGA) for minimizing an Ising Hamiltonian}
	\label{algo:qaga}
\end{algorithm}

\section{Results}
We use randomly generated Ising Hamiltonians for evaluating the performance of the proposed QAGA and compare it with MQC, which is the state-of-the-art technique in the realm of QA~\cite{ayanzadeh2021multi}.
We employ three different types of benchmark problems: 
(a) coefficients drawn uniformly from $\{-1,+1\}$ (binary coefficients); 
(b) coefficients drawn uniformly from $[-1,+1]$ (uniform coefficients); 
and (c) coefficients drawn from the standard normal distribution (normal coefficients).
For our evaluations, we use a D-Wave 2000Q QA. 
Since randomly generated problems are not compatible with the working graph of QAs, we use the minor-embedding heuristic~\cite{cai2014practical} for embedding the arbitrary random graphs to the Chimera topology of the D-Wave 2000Q quantum processors. 
QAGA iteratively contracts variables whose uncertainties are negligible; thus, the Ising Hamiltonian in QAGA has a dynamic structure, and we need to apply the embedding in all iterations of QAGA. Thus, QAGA needs to embed the remaining Ising Hamiltonian to an executable QMI on the target QA. 

We compare the performance of QAGA to QA with ten spin-reversal transforms and longer inter-sample delay (denoted by QA), which is a notably stronger baseline.  
As our second baseline, we apply MQC to the result of the first baseline (denoted by MQC). 
For each QMI, we request 1,000 samples for all methods.

\vspace{-1em}
\subsection{Experiment A}
For every problem in this experiment, we generate a random graph of size 50 with a specified sparsity rate ($s \in \{0.05, 0.25, 0.5, 0.75, 1.0\}$). 
More specifically,  we randomly select edges from a complete graph with $N=50$ nodes where the sparsity rate ${s}$ denotes the probability of selecting edges. 
We then set values of the corresponding biases and couplers randomly.
The uncertainty threshold in QAGA is $\theta=0.0$---i.e., all spins must have the same value so QAGA can fix them for the next stage.
We obtained this threshold empirically by evaluating the performance of QAGA on a small problem set.
We study the sensitivity of QAGA to $\theta$ in the following experiment.

\begin{figure*} [ht]
	\centering
	\includegraphics[scale=0.6]{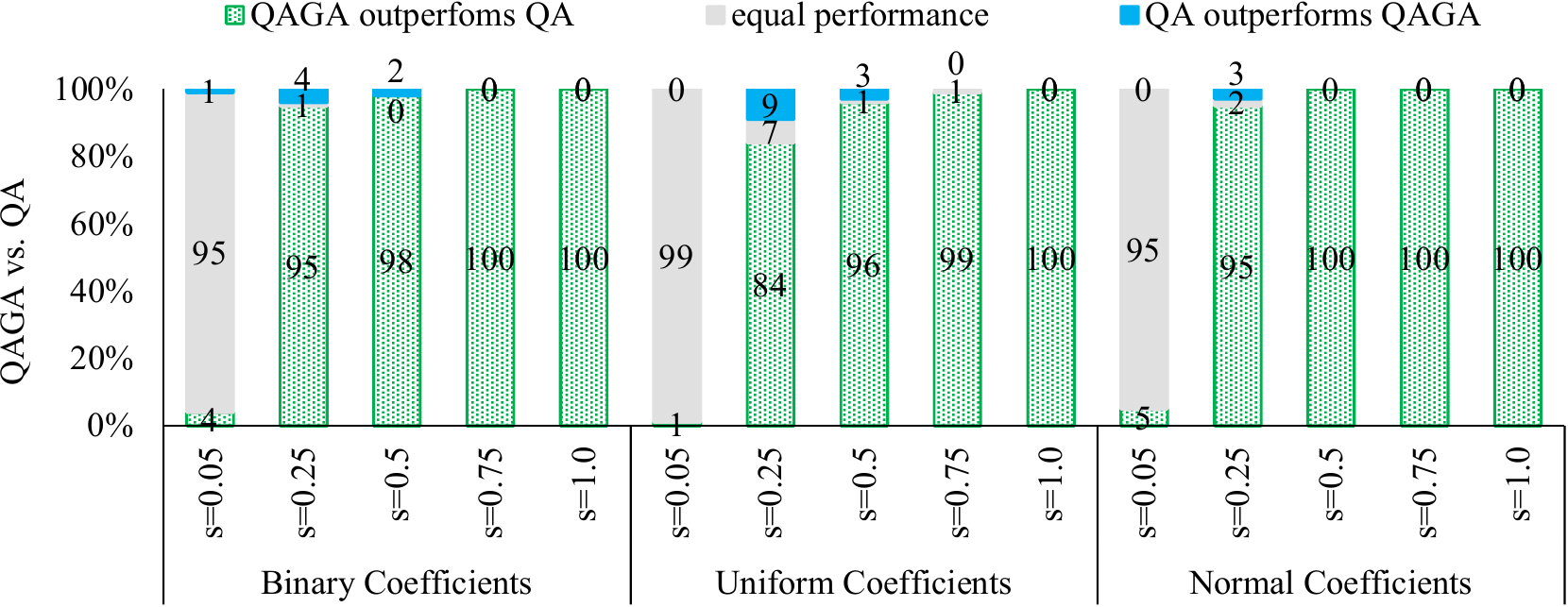}
	\caption{Performance comparison between QAGA and QA (with spin-reversal-transforms and inter-sample delays) in solving 100 random benchmark problems with $N=50$ spin variables and different sparsity rates ${s}$}
	\label{fig:qaga_vs_qa}
\end{figure*}

Figure \ref{fig:qaga_vs_qa} illustrates the performance comparisons between QAGA and QA in minimizing Ising Hamiltonians with binary, uniform and normal coefficients. 
For each case, we generate 100 random problems. 
In this experiment, all randomly generated Ising Hamiltonians include 50 spin variables ($N=50$). 
For each sparsity rate, the corresponding column illustrates the number of times that QA has found a better sample (compared to QAGA), the number of times that QA and QAGA demonstrated the same performance (i.e., best samples from both methods had identical Ising energies), and the number of times that QAGA has outperformed QA. 

\begin{figure*}[ht]
	\centering
	\includegraphics[scale=0.6]{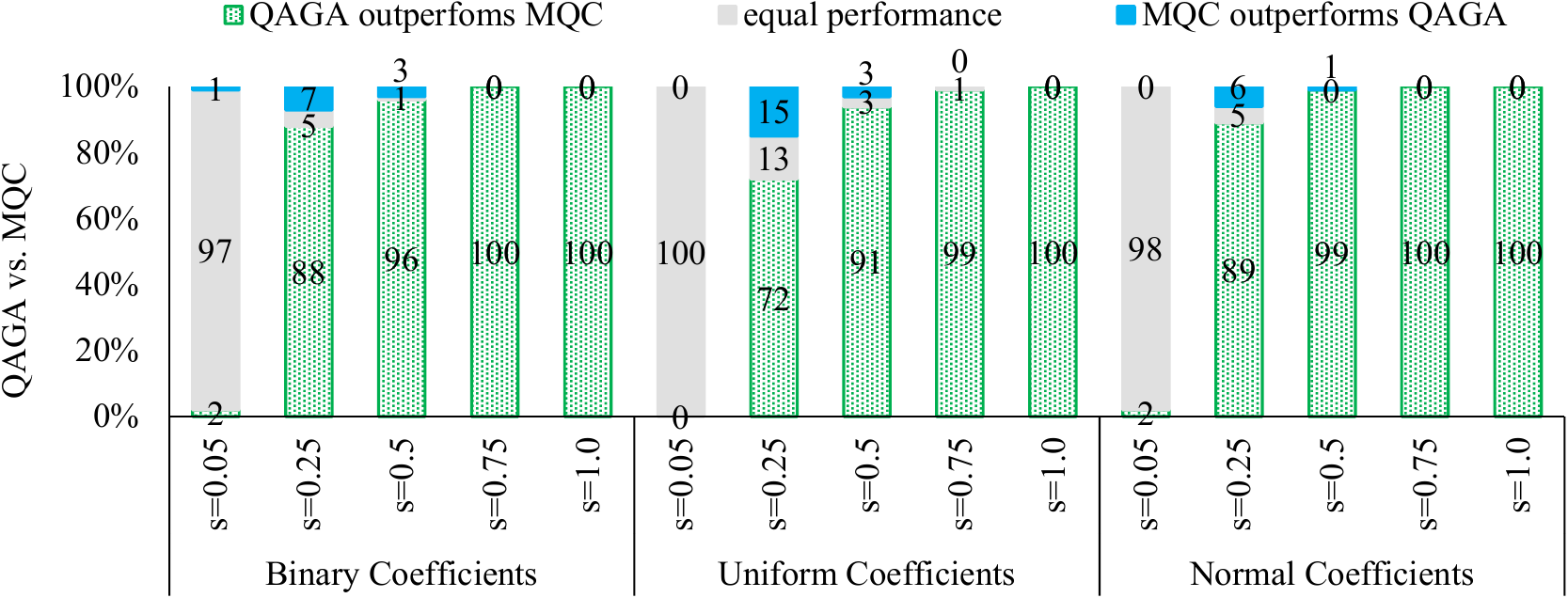}
	\caption{Performance comparison between QAGA and MQC in solving 100 random benchmark problems with $N=50$ spin variables and different sparsity rates ${s}$}
	\label{fig:qaga_vs_mqc}
\end{figure*}

In the same manner, Fig.~\ref{fig:qaga_vs_mqc} illustrates the performance comparisons between QAGA and MQC in minimizing the same 100 randomly generated Ising Hamiltonians with binary, uniform and normal coefficients. 
Note that all arrangements in this experiment (e.g., number of variables, sparsity rate, number of spin-reversal-transforms, etc.) were identical to the previous experiment.
Similar to Fig.~\ref{fig:qaga_vs_qa}, each column in Fig.~\ref{fig:qaga_vs_mqc} represents the number of times that MQC has found a sample with lower energy, MQC and QAGA had similar performance, and QAGA resulted in a sample with a lower Ising energy value. 

\vspace{-1em}
\subsection{Experiment B}
In this experiment, we aim to study the impact of the threshold parameter on the performance of QAGA. 
To this end, we measure the average number of stages (iterations) that QAGA takes to converge. 
Table~\ref{tbl:qaga_num_iterations} illustrates the average number of iterations that QAGA takes to solve 100 random benchmark problems with $N=50$ variables and normal coefficients.

\begin{table}[ht]
	\caption{Average number of iterations for QAGA in solving 100 random benchmark problems with different thresholds }
	\centering
	{%\{resizebox{.95\columnwidth}{!}{	\
	\smallskip\begin{tabular}{c|ccccc}
			&	\multicolumn{5}{c}{sparsity rate (${s}$)}\\ %\cline{2-6}
		$\theta$	&	0.05	&	0.25	&	0.50	&	0.75	&	1.00\\ \cline{2-6}
%--------------------------------------------------------------------------------------------------------
		0.25		&	3.25	&	2.80	&	3.15	&	3.40	&	3.30 \\
		0.15		&	3.60	&	3.40	&	3.35	&	3.45	&	3.65 \\
		0.05		&	4.10	&	4.45	&	4.30	&	3.20	&	4.05 \\
		0.00		&	5.50	&	6.20	&	2.10	&	2.05	&	2.30 \\
\end{tabular}
	}
	\label{tbl:qaga_num_iterations}
\end{table}

Table~\ref{tbl:qaga_num_iterations} reveals that for Chimera-like problems (i.e., sparse problems where $s=0,05$ or $0.25$), when $\theta \to 0$, QAGA takes more iterations (i.e., slower convergence). 
On the other hand, for dense Ising Hamiltonians, where $s \to 1$ (i.e., clique like problems), when $\theta \to 0$, QAGA converges quickly—the maximum number of iterations appears on $\theta \sim 0.1$. 

%\vspace{-1em}
\section{Conclusion}
Quantum annealers are a type of adiabatic quantum computer that can sample from the ground state of Hamiltonians. 
Unfortunately, several technological barriers preclude physical QAs from attaining the ground state of the given problem Hamiltonians. 
%In this study, 
We introduce the notion of {\em quantum-assisted greedy algorithms ({\em QAGA})} that employs QAs for making globally optimum choices at each stage of a greedy algorithm.
QAGA views QAs as a physical process that naturally draws samples from the ground state of Ising Hamiltonians (i.e., a problem-dependent Boltzmann distribution) at cryogenic temperatures.
Combining QAs and greedy algorithms addresses the limitations of both and results in remarkably better solutions, albeit executing multiple QMIs for one problem. 

Our empirical results on a D-Wave 2000Q quantum processor demonstrate that QAGA finds samples with remarkably lower energy values compared to the multi-qubit correction (MQC) method that is the state-of-the-art technique in the realm of QA. 
For sparse problems (i.e., the structure of the problem is close to the Chimera architecture), the performance of the QAGA approaches to MQC.
When the sparsity decreases, however, QAGA shows supremacy in terms of finding samples with lower energy values.

%\subsubsection*{Acknowledgements}
%This research has been supported by NASA grant (\#NNH16ZDA001N-AIST 16-0091), NIH-NIGMS Initiative for Maximizing Student Development Grant (2 R25-GM55036), and the Google Lime scholarship. We would like to thank the D-Wave Systems management team, namely Rene Copeland, for granting access to the D-Wave 2000Q quantum annealer.

% -------------------------------------------------------------------------
\bibliographystyle{IEEEbib}
\bibliography{biblio}

\begin{thebibliography}{10}

\bibitem{preskill2018quantum}
John Preskill,
\newblock ``Quantum computing in the nisq era and beyond,''
\newblock {\em Quantum}, vol. 2, pp. 79, 2018.

\bibitem{huang2021information}
Hsin-Yuan Huang, Richard Kueng, and John Preskill,
\newblock ``Information-theoretic bounds on quantum advantage in machine
  learning,''
\newblock {\em Physical Review Letters}, vol. 126, no. 19, pp. 190505, 2021.

\bibitem{nielsen2010quantum}
Michael~A Nielsen and Isaac~L Chuang,
\newblock {\em Quantum Computation and Quantum Information},
\newblock Cambridge University Press, 2010.

\bibitem{devore1996some}
Ronald~A DeVore and Vladimir~N Temlyakov,
\newblock ``Some remarks on greedy algorithms,''
\newblock {\em Advances in computational Mathematics}, vol. 5, no. 1, pp.
  173--187, 1996.

\bibitem{mousavi2019survey}
Seyedahmad Mousavi, Mohammad Mehdi~Rezaee Taghiabadi, and Ramin Ayanzadeh,
\newblock ``A survey on compressive sensing: Classical results and recent
  advancements,''
\newblock {\em arXiv preprint arXiv:1908.01014}, 2019.

\bibitem{johnson2011quantum}
Mark~W Johnson, Mohammad~HS Amin, Suzanne Gildert, Trevor Lanting, Firas Hamze,
  Neil Dickson, R~Harris, Andrew~J Berkley, Jan Johansson, Paul Bunyk, et~al.,
\newblock ``Quantum annealing with manufactured spins,''
\newblock {\em Nature}, vol. 473, no. 7346, pp. 194, 2011.

\bibitem{mcgeoch2020theory}
Catherine~C McGeoch,
\newblock ``Theory versus practice in annealing-based quantum computing,''
\newblock {\em Theoretical Computer Science}, vol. 816, pp. 169--183, 2020.

\bibitem{ayanzadeh2020leveraging}
Ramin Ayanzadeh,
\newblock {\em {Leveraging Artificial Intelligence to Advance Problem-Solving
  with Quantum Annealers}},
\newblock Ph.D. thesis, U. of Maryland, Baltimore County, 2020.

\bibitem{ayanzadeh2020reinforcement}
Ramin Ayanzadeh, Milton Halem, and Tim Finin,
\newblock ``Reinforcement quantum annealing: A hybrid quantum learning
  automata,''
\newblock {\em Scientific Reports}, vol. 10, no. 1, pp. 1--11, 2020.

\bibitem{ayanzadeh2021multi}
Ramin Ayanzadeh, John Dorband, Milton Halem, and Tim Finin,
\newblock ``Multi-qubit correction for quantum annealers,''
\newblock {\em Scientific Reports}, vol. 11, no. 1, pp. 1--12, 2021.

\bibitem{cai2014practical}
Jun Cai, William~G Macready, and Aidan Roy,
\newblock ``A practical heuristic for finding graph minors,''
\newblock {\em arXiv preprint arXiv:1406.2741}, 2014.

\end{thebibliography}

\end{document}